\documentclass[a4paper,11pt]{article}
\usepackage{amsmath,amssymb,amsfonts,graphicx}

\newcommand{\phibar}{\bar{\phi}}
\newcommand{\tr}{\mathop{\rm tr}\nolimits}
\newcommand{\Slash}[1]{\ooalign{\hfil/\hfil\crcr$#1$}}

\begin{document} 

\begin{flushright} 
DFTT 18/2010\\
RIKEN-TH-196\\
WIS/15/10-OCT-DPPA
\end{flushright} 

\vspace{0.1cm}

\begin{center}
\renewcommand{\thefootnote}{\fnsymbol{footnote}}
    \noindent{\LARGE{
Absence of sign problem 
in two-dimensional ${\cal N}=(2,2)$ super Yang-Mills on lattice
}}\\
    
    \vspace{2cm} 
    
     \noindent{
       Masanori Hanada$^{abc}$\footnote{e-mail:mhanada@u.washington.edu} and 
       Issaku Kanamori$^{dec}$\footnote{e-mail:issaku.kanamori@physik.uni-regensburg.de}}\\
    
    \vspace{1cm}	
    
   $^a$ {\it  Department of Physics, University of Washington,\\
 Seattle, WA 98195-1560, USA} \\ 

    $^{b}${\it Department of Particle Physics and Astrophysics,  
Weizmann Institute of Science,\\ 
Rehovot 76100, Israel} \\

	 $^c$ {\it Theoretical Physics Laboratory,
RIKEN Nishina Center,\\
 Wako, Saitama 351-0198, Japan}\\
 
 $^d$
{\it Institut f\"ur Theoretische Physik, Universit\"at Regensburg,\\ 
D-93040 Regensburg, Germany}\\
 
		$^e$
{\it INFN sezione di Torino, and Dipartimento di Fisica Teorica, 
Universit\`a di Torino,\\
Via P. Giuria 1, 10125 Torino, Italy}

    \vskip 2cm
\end{center}
\setcounter{footnote}{0}

\begin{abstract}
We show that ${\cal N}=(2,2)$ $SU(N)$ super Yang-Mills theory 
on lattice does not have sign problem in the continuum limit, 
that is, under the phase-quenched simulation phase of the determinant 
localizes to 1 and hence the phase-quench approximation becomes 
exact.  
Among several formulations, we study models by Cohen-Kaplan-Katz-Unsal
 (CKKU) and by Sugino.  
We confirm that the sign problem is absent in both models and 
that they converge to the identical continuum limit 
\emph{without fine tuning}.  
We provide a simple explanation why previous works by other authors, 
which claim an existence of the sign problem, 
do not capture the continuum physics.

\end{abstract} 

\newpage
\section{Introduction}
Supersymmetric Yang-Mills theory (SYM) attracts broad interests 
as a candidate of the physics beyond the standard model \cite{Martin97}. 
It is also a promising candidate for the nonperturbative formulation of 
the superstring theory \cite{BFSS96,IKKT96,DVV97,Maldacena97,IMSY98}. 
Although it is important to study its nonperturbative properties for both 
applications, and one of the most successful nonperturbative approaches 
for gauge theory is the lattice simulation, 
however, notorious difficulty of the lattice SYM prevented 
it for long time. Sometime ago there appeared a breakthrough for
two-dimensional theories 
of extended supersymmetries 
\cite{CKKU03,Sugino04,2d other formulations,KU_16SUSY,D'Adda:2005zk},  
(see also \cite{Catterall:2009it} for a recent review);
the correct continuum limit is 
obtained without parameter fine tunings for most of these models, 
at least at perturbative level.\footnote{
There is a lattice formulation of 3d maximally supersymmetric Yang-Mills 
without fine tuning \cite{KU_16SUSY}. 
For 4d $\mathcal{N}=1$ pure supersymmetric Yang-Mills, the chiral
symmetry guarantees supersymmetric continuum limit.
Recent results are found in \cite{4d-n1-SYM}.
}
Furthermore it turned out that by combining two-dimensional lattice with
matrix model technique (fuzzy sphere) \cite{Maldacena:2002rb}, 
4d ${\cal N}=4$ theory is constructed 
without relying on the parameter fine tunings to all order in
perturbation theory \cite{HMS10}. (For other very elegant formulation 
in the planar limit, which preserves 16 supersymmetries manifestly, 
see \cite{Ishii:2008ib}.
This method is applicable to various kind of theories for which  
lattice and/or fuzzy sphere technique are not applicable \cite{Hanada:2009hd}.) 
These formulations provide robust ways to test the $AdS_5$/$CFT_4$
correspondence, the gauge/gravity duality 
\cite{Maldacena97,IMSY98} and matrix string conjecture \cite{DVV97} 
at nonperturbative level. 
In one dimension, such a program has been pursued extensively 
by using a non-lattice method \cite{HNT07} 
and the gauge/gravity duality \cite{Maldacena97,IMSY98} has been confirmed 
very precisely \cite{AHNT07,HMNT08,HHNT08,HNSY09, HNSY10}, 
including the stringy $\alpha'$ corrections \cite{HHNT08,HNSY10}. 
The lattice simulations are also applicable for this system and 
qualitatively consistent results have been obtained \cite{CW07}. 
We can expect study of two and four dimensional theories will provide
even richer insights; 
in two dimensions simulations in this context are already ongoing
\cite{HK09,Catterall:2010fx, Catterall:2010ya} 
and it is urgent to establish the validity of the lattice models at
nonperturbative level by detailed simulations.

However there is a possible obstacle for simulations: supersymmetric theories 
with eight and sixteen supersymmetries suffer from 
the sign problem \cite{KNS98}\footnote{
In the case of maximally supersymmetric matrix quantum mechanics,
agreement with the dual gravity
prescription has been observed by ignoring the phase of the Pfaffian,
even when the sign fluctuates violently \cite{HMNT08,HNSY09}. It has
also been observed that the Pfaffian is almost real positive for 
$SU(2)$ \cite{HNSY10}. It would be nice to understand why it happens, 
but it is out of scope of the present paper.}
\footnote{
Even if there is a sign problem, measurement of the sign factor
 itself is interesting because it is related
to the Witten index with a suitable normalization \cite{Kanamori:2010gw}.
}.
On the other hand, with four supersymmetries (i.e. 
4d ${\cal N}=1$ pure SYM and its dimensional reductions), 
there is no sign problem. It can easily be seen as follows. In Weyl
notation, with an appropriate choice of the gamma matrices, the Dirac
operator is written as
\begin{eqnarray}
 D
\equiv
i\sigma^\mu D_\mu, 
\end{eqnarray}
where $\sigma^0=-i\textbf{1}_2$ and $\sigma^i(i=1,2,3)$ are Pauli matrices. 
By using $\sigma^2(i\sigma^\mu)\sigma^2=(i\sigma^\mu)^\ast$  
and the fact that $D_\mu$ is real in adjoint representation, we obtain 
\begin{eqnarray}
 \sigma^2 \Slash{D}\sigma^2
  =
  D^\ast.  
\end{eqnarray}
Therefore, if $\varphi$ is an eigenvector corresponding to an eigenvalue $\lambda$, 
$\sigma^2\varphi^\ast$ is also an eigenvector, with eigenvalue $\lambda^\ast$.
They are linearly independent and eigenvalues appear in a pair
$(\lambda,\lambda^\ast)$.
This assures the positivity of the determinant after removing
$\lambda=0$ modes.

At discretized level, positivity of the determinant can be lost. 
In zero dimension (matrix model) \cite{AABHN00}, there is no sign problem, 
because there is no need for the regularization. 
In 1d theory and fuzzy sphere construction 
of three and four dimensional theories \cite{Ishii:2008ib}, 
by using the momentum cutoff prescription
\cite{HNT07} the determinant is positive even at discretized level
\cite{HMM09}. 

In lattice constructions of two-dimensional SYM, the determinant is 
in general complex at discretized level.  
In Sugino's model \cite{Sugino04} for two-dimensional theory, however, 
the sign problem disappears as one approaches to 
the continuum~\cite{Suzuki07,Kanamori09};
that is, if one performs the phase-quenched simulation, distribution of 
the phase factor of the determinant\footnote{
Strictly speaking, in Sugino model fermions are Majorana and hence we
calculate the Pfaffian.
} peaks at 1 in the continuum limit. 
Therefore the phase quench approximation becomes exact at continuum.  
In this case, in addition to the absence of sign, the agreements with 
analytic calculations in small volume region have been observed as well
\cite{HK09} by using techniques developed in \cite{Azeyanagi:2009zf}. 
Therefore the absence of the sign is the property of the correct
continuum limit, as expected.
Numerical studies of this model can also be found in \cite{KSS07,KS08}.

On the other hand, for other two-dimensional lattice models 
which are supposed to have the same continuum limit, 
an \emph{existence} of the sign problem has been reported \cite{Giedt03,Catterall08}. 
In Cohen-Kaplan-Katz-Unsal (CKKU) model, which is equivalent to 
a model by D'Adda et.al.~\cite{D'Adda:2005zk} with a specific choice of 
parameters, Giedt reported that the determinant has a complex phase, 
and the phase fluctuates violently if one chooses random lattice
configurations \cite{Giedt03}. 
However the importance sampling has not been performed in
\cite{Giedt03}, and hence this result has nothing to do with the
continuum limit as the author remarked correctly.
This model was studied later with importance sampling in
\cite{Catterall08}\footnote{
In \cite{Catterall08}, the bosonic fields are defined 
as $e^{z}$ where $z$ is a complex field, and thus a natural way of
extracting physical quantities is different from the original CKKU model. 
For $SU(N)$ gauge group as in \cite{Catterall08}, this definition is
different from a general complex field originally defined in
\cite{CKKU03}.  
We would like to thank S.~Catterall for detailed explanations on
his work.}, 
where the sign problem was reported as well.
However, it is not clear whether it is a property of the continuum,
because they could not evaluate physical quantities because of the
``sign problem'' and hence could not estimate how close to the continuum
limit they have reached.

In this paper, we resolve the confusion mentioned above. 
We show the absence of sign problem in the Sugino model and 
the CKKU model  
in the continuum limit. (In Sugino model, the absence of sign 
has been explicitly reported for $SU(2)$ theory 
in \cite{Suzuki07,Kanamori09}.  
For $N>2$ theory, we briefly checked but have not mentioned 
it in \cite{HK09}, because our emphasis was put 
on other physical quantities. 
In this paper we show the detail for $SU(N)$ with $N>2$, 
together with new data for $SU(2)$.)  
The action in the continuum is obtained from 4d ${\cal N}=1$ SYM through
the dimensional reduction, and is given by 
\begin{align}
S
&=
\frac{N}{\lambda}\int_0^{L_x} dx\int_0^{L_y}dy
\ Tr
\left\{
\frac{1}{4}F_{\mu\nu}^2
+
\frac{1}{2}(D_\mu X_i)^2
-
\frac{1}{4}[X_i,X_j]^2
-
\frac{1}{2}\bar{\psi}\Gamma^\mu D_\mu\psi
-
\frac{i}{2}\bar{\psi}\Gamma^i [X_i,\psi]
\right\}, 
\end{align}
where $\mu$ and $\nu$ run $x$ and $y$, $i$ and $j$ run $1$ and $2$, 
and $\Gamma^I=(\Gamma^\mu,\Gamma^i)$ are gamma matrices in four dimensions.   
$X_i$ are $N\times N$ hermitian matrices, $\psi_\alpha$ are $N\times N$ 
fermionic matrices with a Majorana index $\alpha$ 
and the covariant derivative 
is given by $D_\mu=\partial_\mu-i[A_\mu,\ \cdot\ ]$. 
The only parameters of the model are 
the size of circles $L_x$ and $L_y$. (Note that the coupling constant 
can be absorbed by redefining the fields and coordinates. 
Therefore we take the 't Hooft coupling $\lambda$ to be $1$.
Then the strong coupling corresponds to the large volume.) 
We study this system by using two lattice models (CKKU and Sugino) 
numerically and show
the absence of the sign problem.
We evaluate expectation values of some physical quantities and see 
that the results show nice agreements.
Note that small volume behavior of the Sugino model is consistent with 
known analytic estimates \cite{HK09}. 

One obstacle for the simulation is the existence of the flat direction, 
along which two scalar fields $X_1$ and $X_2$ commute. 
In contrary to a theory on ${\mathbb R}^{1,3}$, there is no
superselection of the moduli parameter in this case. That is,
eigenvalues of scalars are determined dynamically.  
Therefore, some mechanism which restrict eigenvalues to a finite distribution 
is necessary for the stable simulation. 
In addition, to obtain an interesting dynamical system,
having a (small) finite region for the eigenvalues is important as well;
if the eigenvalues of the scalar spread so large, the theory would
run into the abelian phase, which is just a free theory\footnote{
Which phase is preferred is in fact a dynamical question.
At large-$N$, the flat direction is lifted and the system
stays non-abelian phase; see \cite{HK09}. 
This phase is an analogue of the black 1-brane solution in type IIB supergravity. 
}. 
In this work, we introduce soft SUSY-breaking mass to scalar fields 
\begin{equation}
 \mu^2 N\int d^2x\sum_{i=1,2} Tr X_i^2, 
\end{equation}
so that the flat direction is lifted%
\footnote{
For 8 and 16 SUSY models, there exists SUSY-preserving mass deformation
\cite{HMS10}. 
}.
It is crucial to control the flat direction for various reasons. 
We have just mentioned two of them 
--- stability of the simulation and interesting non-abelian phase. 
There is one more; in order to guarantee the correct continuum limit,
the eigenvalue must be smaller than the cut off scale $\sim 1/a$.
Especially in the CKKU model, 
we can decompose the bosonic field to appear scalars $X$
as a log of positive Hermitian variables $H$ \cite{Unsal:2005yh},
\begin{equation}
 H=\exp (aX), 
\end{equation}
where $a$ is the lattice spacing. In order to obtain the tree level action, 
one has to assume $aX\ll 1$, expand it in powers of $aX$ and neglect
higher order terms.  
Therefore, unless $aX\ll 1$, one cannot get to the continuum limit even
at tree level.  
Actually it turns out that one of previous works, whose conclusion contradicts 
with ours, does not satisfy this condition.

This paper is organized as follows. In \S~\ref{sec:CKKU model} we study 
the CKKU model. We introduce the model in \S~\ref{sec:CKKU action} and then 
consider a structure of the Dirac operator in \S~\ref{sec:CKKU_zeromode}. 
Then we show the absence of the sign problem in \S~\ref{sec:CKKU_absence_sign}. 
In \S~\ref{sec:Sugino model} we show the absence of the sign problem in
Sugino model. 
Then in \S~\ref{sec:CKKU_SUGINO_COMPARISON} 
we confirm that two models (CKKU and Sugino) 
converge to the same continuum limit. 
To our best knowledge, this is the first result from the CKKU model in the
continuum limit and shows in fact we can take a supersymmetric continuum
limit without any fine tunings.
In \S~\ref{subsec:Giedt} we explain 
why previous works by other authors fail to capture the continuum physics.

\section{CKKU model}\label{sec:CKKU model}
In this section we study the CKKU model. 
In \S~\ref{sec:CKKU action} we introduce the model. 
Then in \S~\ref{sec:CKKU_zeromode} we discuss the structure of light modes 
in the model, which is crucial for the analysis of the sign problem  
shown in \S~\ref{sec:CKKU_absence_sign}. 

\subsection{The model}\label{sec:CKKU action}
Here we consider the $U(N)$ gauge group. Note that the $U(1)$ part 
is decoupled in the continuum limit 
and hence the physics is the same as $SU(N)$ theory.  
The action is given by 
\begin{eqnarray}
S
&=&
Na^2\sum_{\vec{n}}Tr\Biggl\{
\frac{1}{2}\left(
\bar{x}_{\vec{n}-\hat{x}} x_{\vec{n}-\hat{x}}
-
x_{\vec{n}}\bar{x}_{\vec{n}}
+
\bar{y}_{\vec{n}-\hat{y}} y_{\vec{n}-\hat{y}}
-
y_{\vec{n}}\bar{y}_{\vec{n}}
\right)^2
+
2|x_{\vec{n}}y_{\vec{n}+\hat{x}}-y_{\vec{n}}x_{\vec{n}+\hat{y}}|^2
\nonumber\\
& &
\qquad
+
\sqrt{2}\left(
\alpha_{\vec{n}}\bar{x}_{\vec{n}}\lambda_{\vec{n}}
-
\alpha_{\vec{n}-\hat{x}}\lambda_{\vec{n}}\bar{x}_{\vec{n}-\hat{x}}
\right)
+
\sqrt{2}\left(
\beta_{\vec{n}}\bar{y}_{\vec{n}}\lambda_{\vec{n}}
-
\beta_{\vec{n}-\hat{y}}\lambda_{\vec{n}}\bar{y}_{\vec{n}-\hat{y}}
\right)
\nonumber\\
& &
\qquad
-
\sqrt{2}\left(
\alpha_{\vec{n}}y_{\vec{n}+\hat{x}}\xi_{\vec{n}}
-
\alpha_{\vec{n}+\hat{y}}\xi_{\vec{n}} y_{\vec{n}}
\right)
+
\sqrt{2}\left(
\beta_{\vec{n}}x_{\vec{n}+\hat{y}}\xi_{\vec{n}}
-
\beta_{\vec{n}+\hat{x}}\xi_{\vec{n}} x_{\vec{n}}
\right)
\nonumber\\
& &
\qquad
+
a^2\mu^2\left(
x_{\vec{n}}\bar{x}_{\vec{n}}
-
\frac{1}{2a^2}
\right)^2
+
a^2\mu^2\left(
y_{\vec{n}}\bar{y}_{\vec{n}}
-
\frac{1}{2a^2}
\right)^2
\Biggl\}
\nonumber\\
& &
\qquad
+
Na^2\sum_{\vec{n}}
\Biggl\{
a^2\nu^2
\left|
\frac{Tr(x_{\vec{n}}\bar{x}_{\vec{n}})}{N}
-
\frac{1}{2a^2}
\right|^2
+
a^2\nu^2
\left|
\frac{Tr(y_{\vec{n}}\bar{y}_{\vec{n}})}{N}
-
\frac{1}{2a^2}
\right|^2
\Biggl\}. 
\end{eqnarray}
Here $x,y$ are $N\times N$ complex matrices and 
$\alpha,\beta,\lambda,\xi$ are $N\times N$ complex Grassmanian matrices. 
They are related to the fields in the continuum by  
\begin{eqnarray}
& &
x
=
\frac{1}{a\sqrt{2}}+\frac{X_1+iA_1}{\sqrt{2}}, 
\qquad
y
=
\frac{1}{a\sqrt{2}}+\frac{X_2+iA_2}{\sqrt{2}},
\label{identification_boson_naive}
\\
& &
\psi=
\left(
\begin{array}{c}
\lambda\\
\xi
\end{array}
\right), 
\qquad
\bar{\psi}
=
i(\alpha,\beta). 
\end{eqnarray}
The fermion is in the Weyl representation and is complex. 
Hence we study the determinant of the Dirac operator rather than the Pfaffian. 
The Dirac operator $D$ is obtained by writing the fermion part as 
$\bar{\psi}_{ji \alpha x}D_{ij\alpha x,kl\beta y}\psi_{kl\beta y}$,
where suffixes $i,j,k,l$ refer to color, $\alpha,\beta$ to spinor and
$x,y$ to coordinate.  

It is convenient to introduce (semi-)compact decomposition
of the bosonic fields \cite{Unsal:2005yh} 
\begin{eqnarray}
x
=
\frac{1}{\sqrt{2}a}U_1 H_1, 
\qquad
y
=
\frac{1}{\sqrt{2}a}U_2 H_2, 
\end{eqnarray}
where $U_i$ are unitary, $H_i$ are Hermitian and positive definite, 
and 
\begin{eqnarray}
U_i=\exp(iaA_i), 
\qquad
H_i=\exp(aX_i). 
\end{eqnarray}
From $x$ and $y$, $H_i$ can be obtained as 
\begin{eqnarray}
H_1=\sqrt{2a^2x^\dagger x}, 
\qquad
H_2=\sqrt{2a^2y^\dagger y}. 
\end{eqnarray}

A mass parameter $\mu$ gives mass to $U(N)$ scalar fields. 
The last two terms are not present in the original proposal\footnote{
We would like to thank O.~Aharony for suggesting 
the use of $U(1)$ mass term. }; 
it gives mass only to $U(1)$ part of the scalar.
As observed in \cite{AHNT07,HK09}, at large volume and/or with periodic
boundary condition for the fermion, 
flat direction in $SU(N)$ sector is dynamically lifted and nonabelian
phase (i.e. bound state of scalar eigenvalues) becomes meta-stable. 
(It becomes stabler as $N$ increases.) 
On the other hand $U(1)$ flat direction is never lifted, and in the CKKU model, 
it can destroy the lattice structure. But to stabilize this $U(1)$ flat
direction we do not have to turn on the $U(N)$ mass $\mu$; 
the $U(1)$ mass $\nu$ is fine enough. 
Given that the $U(1)$ sector is free and decouples from the dynamics, 
nonzero value of $\nu$ does not affect the supersymmetry i
n the $SU(N)$ sector in the continuum limit. 
We explicitly confirm this statement numerically (See 
Figs.~\ref{fig:wilson_u1massdep},
\ref{fig:su_scalar_u1massdep}, \ref{fig:un_det_u1massdep},
\ref{fig:cos_u1massdep}).
Note that this decoupling contains a delicate issue. 
At lattice level, the $U(1)$ and
$SU(N)$ secters are not completely decoupled. 
In order to stabilize the lattice structure, 
the heavy $U(1)$ mass is suitable.
However, if the $U(1)$ mass is too large, 
the SUSY breaking effect in the $U(1)$ sector becomes large
and it might be mediated through 
the lattice artifact to the $SU(N)$ sector.
Our numerical results support the
decoupling in a wide range of the $U(1)$ mass. 
It would be nice if the decoupling could be
explained analytically. 
We will show that with non-zero value of $\nu$ the obtained
values of the observables are the same as those from $SU(N)$ 
Sugino model (see sec.~\ref{sec:CKKU_SUGINO_COMPARISON}),
which justifies the treatment of the $U(1)$ mass term. 
At finite volume and finite $N$ we need nonzero $\mu$ to completely remove 
the instability, 
but at large-$N$ we can take $\mu=0$ and $\nu\neq 0$ so that the
supersymmetry in the $SU(N)$ sector 
is fully restored already at finite volume.   
It enables one to study interesting finite-volume physics like black hole/black 
string phase transition \cite{Gregory:1993vy,Aharony:2004ig}. 

\subsection{Structure of light modes}\label{sec:CKKU_zeromode}
At the classical vacuum of the $U(N)$ theory, the Dirac operator has $2N^2$ 
fermion zero-modes, which correspond to zero-momentum.  
Apart from the classical vacuum, $2(N^2-1)$ of them are lifted and there remain 
two zero-modes which correspond to the $U(1)$ part. 
At a discretized level, because of the special property of the CKKU model, 
only one of them is exactly zero \cite{Giedt03}.
The other approaches to zero in the
continuum limit. Let us call it the \emph{pseudo zero-mode}.   
 
In our simulation, the phase quenched ensemble with $\det\sqrt{MM^\dagger}$
is used, where $M=iD$.
To avoid the exact zero-mode, we add a regulator term to $MM^\dagger$ in the
simulation,  
\begin{eqnarray}
MM^\dagger\to MM^\dagger+\epsilon\textbf{1}.  
\end{eqnarray}  
In practice $\epsilon$ is fixed to be a small enough value
($\epsilon\sim 10^{-6}$) 
compared to all nonzero eigenvalues of $M M^\dagger$.

When we calculate the determinant, we remove the exact zero-mode and 
the pseudo zero-mode by hand. (In practice we remove one exact 
zero eigenvalue and smallest nonzero eigenvalue of the Dirac operator.) 
As we will see, removal of the pseudo zero-mode is crucial to establishing 
the positivity of the determinant. The reason is simple -- 
in the continuum, the Dirac operator has positive determinant because 
its eigenvalues form pairs $(\lambda,\lambda^\ast)$ \emph{after}
 removing zero-modes. 
The reason why we \emph{can} remove this pseudo zero mode is clear; 
it will decouple from the dynamics in any case.
The corresponding zero-modes which should be removed in the lattice
simulation are the exact zero-mode and the pseudo zero-mode.
If the lattice artifact to zero-modes had a pair structure which keeps
the positivity of the determinant we would not need to remove these two modes. 
However, phase of the pseudo zero-mode fluctuates violently 
and it dominates the fluctuation of the phase of the determinant 
because pseudo zero-mode does not appear in a pair; without removing it 
the phase of the determinant becomes completely random, just because the phase of 
the pseudo zero-mode is random. 

We numerically calculated the eigenvalues and the determinant of $iD$.
We observed that at very small lattice spacing,
the other $2(N^2-1)$ light modes have a pair structure
$(\lambda,-\lambda^\ast)$, which is exactly expected from the continuum
argument.

\subsection{Absence of the sign problem}\label{sec:CKKU_absence_sign}
\begin{figure}
\hfil
\includegraphics[width=0.49\linewidth]{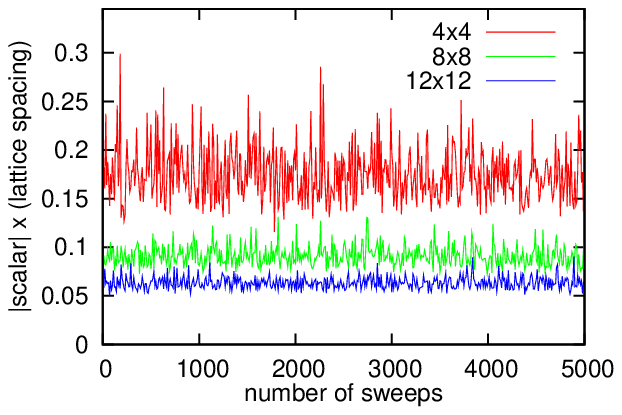}
\hfil
\includegraphics[width=0.49\linewidth]{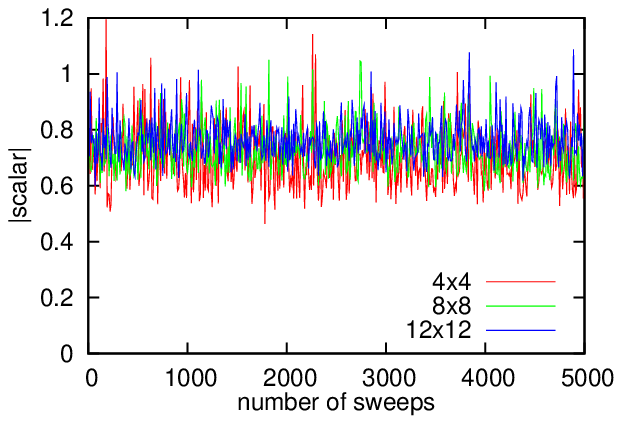}
  \caption{ 
[CKKU] History of the average scalar extent,
  in the lattice unit (left) and in the physical unit (right).
 At each sweep the average over $x$ and $i$ of $\sqrt{Tr(aX_i(x))^2/N}$ 
(a contraction w.r.t. $i$ 
is {\it not} taken), where $aX_i\equiv\log H_i$,   
is plotted.  
Clear convergence as $\sim a$ in the left panel, 
which corresponds to fixed physical extent of the scalar in the right
 panel, can be seen. 
 The physical volume is $L=1.0$, the $U(N)$ mass is
 $\mu=1.0$ and the $U(1)$ mass is $\nu=5.0$. The gauge group is $U(2)$.}
 \label{fig:U2_Vol=100_M100U000scalar}
\end{figure}

\begin{figure}
 \hfil
\includegraphics[width=0.49\linewidth]{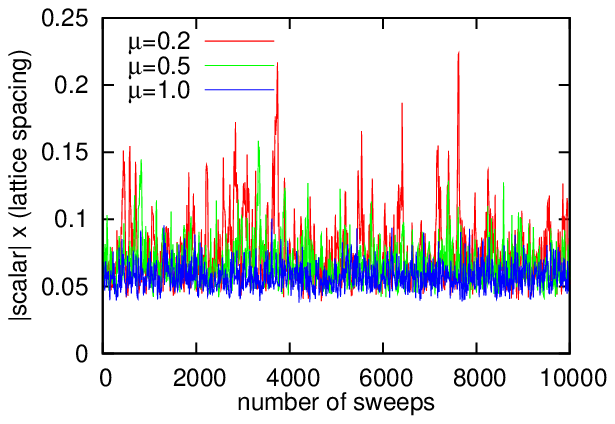}
 \hfil
\includegraphics[width=0.49\linewidth]{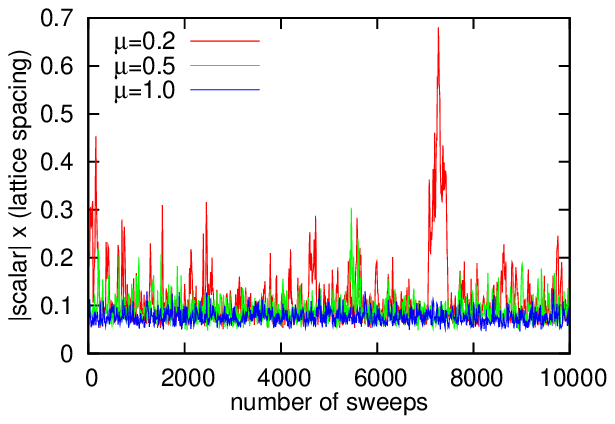}

  \caption{ [CKKU] History of the average scalar extent 
in the $U(2)$ theory at $L=0.5$, $\nu=5.0$ and $\mu=1.0, 0.5$ and $0.2$. 
 At each sweep the average of $\sqrt{Tr (aX_i(x))^2/N}$ is plotted. 
 (left) $8\times 8$ lattice, even at $\mu= 0.2$ the flat direction 
is under control. 
 (right) $6\times 6$ lattice, the flat direction appears around
 $\mu=0.2$.
 Note the difference of the scales between the panels.
}
\label{fig:U2_Vol=050_U10_scalar_M_dep}
\end{figure}

\begin{figure}
 \hfil
 \includegraphics{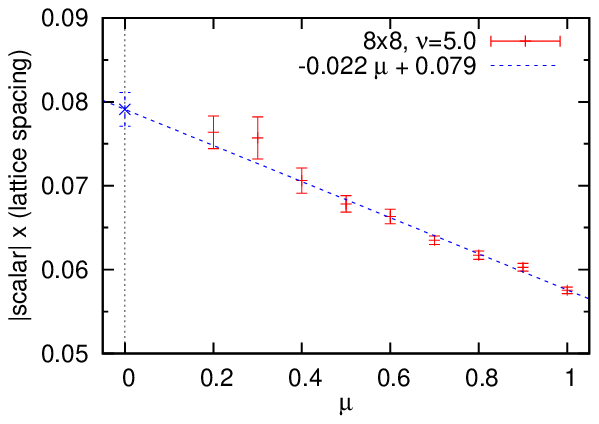}
  \caption{ [CKKU] Mass dependence of 
$\left\langle \sqrt{\frac{1}{N} TrX_i^2(x)}\right\rangle\times
({\rm lattice\ spacing})$ 
in the $U(2)$ CKKU model.  
As in Figs.~\ref{fig:U2_Vol=100_M100U000scalar} and
\ref{fig:U2_Vol=050_U10_scalar_M_dep},
a contraction w.r.t. $i$ is not taken and
an average over $i$ and $x$ is taken.
The lattice size is $8\times 8$, the physical volume is $0.50\times 0.50$ and 
the $U(1)$ mass is $\nu=5.0$. 
}
 \label{fig:CKKU_scalar_extent}
\end{figure}

As we have explained, it is important to control the flat
direction in order to study continuum physics.  
In Fig.~\ref{fig:U2_Vol=100_M100U000scalar} 
we show histories of the extent of the scalar fields 
in the lattice unit (left) and in the physical unit (right).  
When the extent is close to the cutoff scale, 1 in the lattice unit, 
the simulation is not reliable.  
As we can see from the left panel of
Fig.~\ref{fig:U2_Vol=100_M100U000scalar}, at the physical volume $L=1.0$,
the $U(N)$ mass $\mu=1.0$ and the $U(1)$ mass $\nu=5.0$, 
scalars take sufficiently small values,  
which shows that the flat direction is well under control to 
guarantee the correct continuum limit.  
Thus there is no instability caused by a lattice artifact.
Moreover the scalar
fields stay finite in physical unit (the right panel).  That is, this lattice 
model describes a system without the flat direction in the continuum limit.
The obtained continuum system has no instability along the flat direction.
It is expected of course, because we added a mass term. 
Note that if the scalar converged in the lattice unit but diverged in
physical unit we would obtain a continuum limit with flat direction, 
which is just a free theory in the Abelian phase.
As the mass is decreased, the flat direction emerges gradually. 
In Fig.~\ref{fig:U2_Vol=050_U10_scalar_M_dep} we show
the histories at $L=0.5$ with a few values of $\mu$.
With $8\times 8$ ($a=0.0625$) lattice the fluctuation is not violent 
even at $\mu=0.2$ (the left panel). 
As one can easily imagine, with smaller lattice the instability 
-- caused by the lattice artifact -- appears more easily; see  
the right panel. 
This plot uses the same physical parameters as the left panel, 
but a smaller lattice.  
Note that for the smaller lattice (i.e. larger lattice spacing)
more spikes appear at $\mu=0.2$, which is a signal of
the instability.
The reason is obvious -- 
with smaller $\mu$ the extent of the scalar in physical unit is larger, 
and to make it to be small in lattice unit lattice spacing must be smaller. 
By extrapolating to $\mu=0$ by using data from stable region, we obtain 
a finite extent of the scalar (Fig.~\ref{fig:CKKU_scalar_extent};
here we have assumed a simple linear extrapolation, based on the
obtained plot). 
Therefore we can expect the phase we are looking at is smoothly connected to 
the meta-stable non-abelian phase \cite{HNT07,AHNT07,HK09}.   
 
In the following, for all data we show, we have confirmed the simulation 
does not run away to the flat direction. 

Now we are ready to study the sign problem.
As discussed in detail in Appendix \ref{appendix:eigenmodes},
the determinant of $iD$ has a sign $(-1)^{N-1}$, at least
for the constant configurations.
Therefore, throughout this section, we multiply $(-1)^{N-1}$ to the phase factor
of the determinant so that it localizes around $+1$.

In short, what happens both in the CKKU and Sugino is  
\begin{itemize}
\item
For fixed lattice spacing, the phase fluctuation becomes smaller 
at smaller volume and/or smaller $N$. 

\item
For fixed $N$ and fixed volume, the phase disappears in the continuum limit 
(small lattice spacing). 

\end{itemize}

\begin{figure}
 \hfil
 \includegraphics[width=0.49\linewidth]{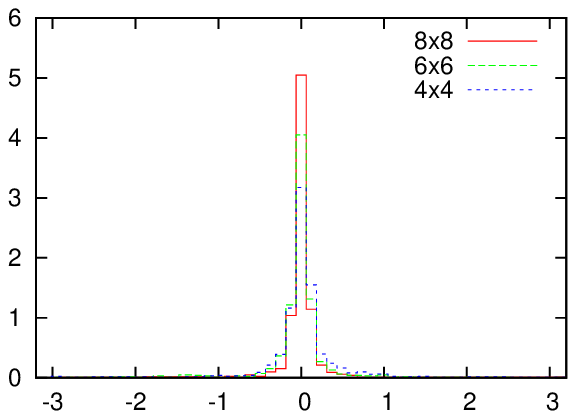}
 \hfil
 \includegraphics[width=0.49\linewidth]{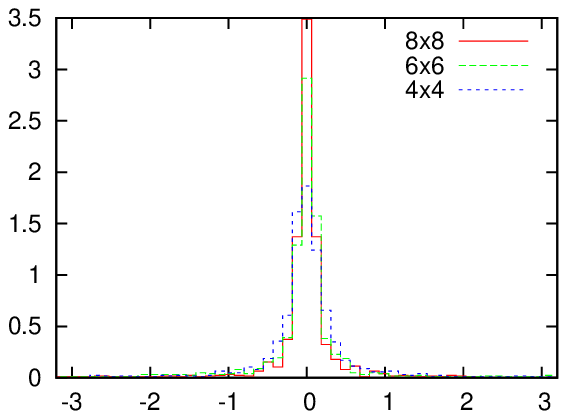}
  \caption{[CKKU] Distribution of the argument of the determinant 
 in $U(2)$ theory. 
 The $U(N)$ mass is $\mu=1.0$ and the $U(1)$ mass is $\nu=5.0$.
 The physical volume is $L=0.5$ (left) and $L=0.75$ (right). 
 }
 \label{fig:u2_det_m1.0}
\end{figure}

Let us start with the $U(2)$ theory. In Fig.~\ref{fig:u2_det_m1.0}
we have shown how the distributions of the argument of the determinant
peaks to $0$.  In each panel, the physical volume is fixed 
and the number of sites is changed. 
There is a clear tendency that the peak becomes sharper as one goes closer 
to the continuum. 

\begin{figure}
 \hfil
 \includegraphics[width=0.49\linewidth]{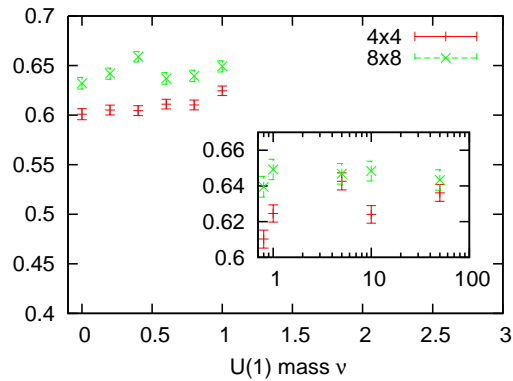}

 \caption{[CKKU] The $U(1)$ mass dependence of the Wilson loop in the $U(2)$
 CKKU model.
 The $U(N)$ mass is fixed to $\mu=1.0$ and the physical volume is fixed to
 $1.0\times 1.0$.
 }
 \label{fig:wilson_u1massdep}
\end{figure}

\begin{figure}
 \hfil
 \includegraphics[width=0.49\linewidth]{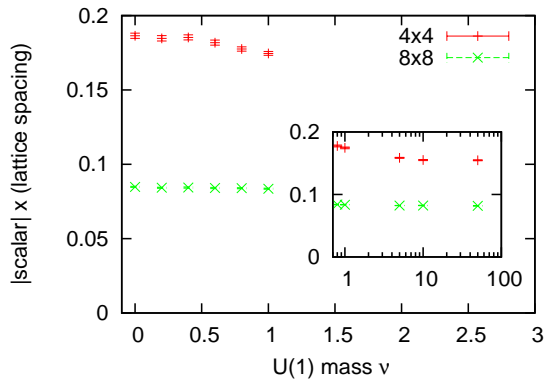}

 \caption{[CKKU] The $U(1)$ mass dependence of 
$\left\langle \sqrt{\frac{1}{N} TrX_i^2(x)}\right\rangle\times
({\rm lattice\ spacing})$ 
in the $U(2)$ CKKU model.  
As in Figs.~\ref{fig:U2_Vol=100_M100U000scalar} and others, 
a contraction w.r.t. $i$ is not taken and
an average over $i$ and $x$ is taken.
The $U(N)$ mass is fixed to $\mu=1.0$ and the physical volume is fixed to
 $1.0\times 1.0$.
 }
\label{fig:su_scalar_u1massdep}
\end{figure}

\begin{figure}
 \hfil
 \includegraphics[width=0.49\linewidth]{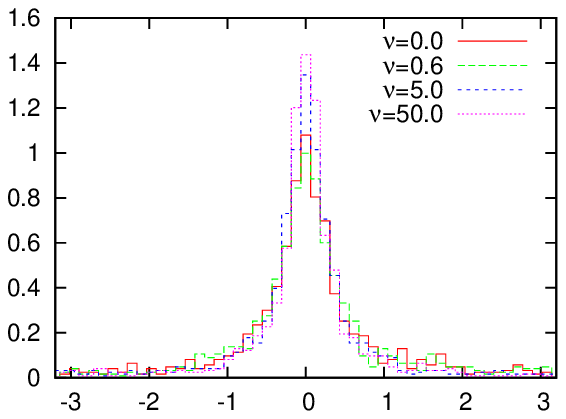}
 \hfil
 \includegraphics[width=0.49\linewidth]{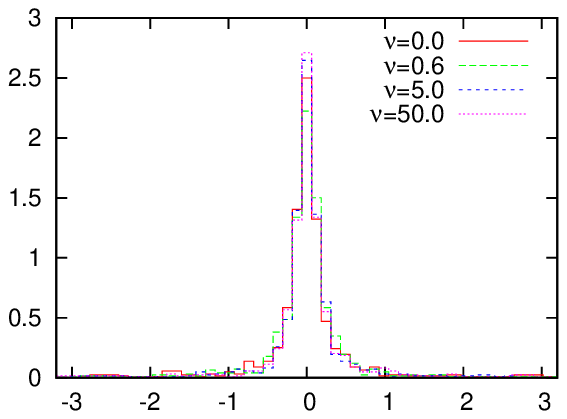}

 \caption{[CKKU] The distribution of the argument of the determinant in
the $U(2)$ CKKU model, with various values of the $U(1)$ mass $\nu$.
 The left panel is for $4\times 4$ lattice and the right is for 
$8\times 8$ lattice.
 The $U(N)$ mass is fixed to $\mu=1.0$ and the physical volume is fixed to
 $1.0\times 1.0$.  A factor $(-1)^{N-1}$ is multiplied.
 }
 \label{fig:un_det_u1massdep}
\end{figure}

\begin{figure}
 \hfil
 \includegraphics[width=0.49\linewidth]{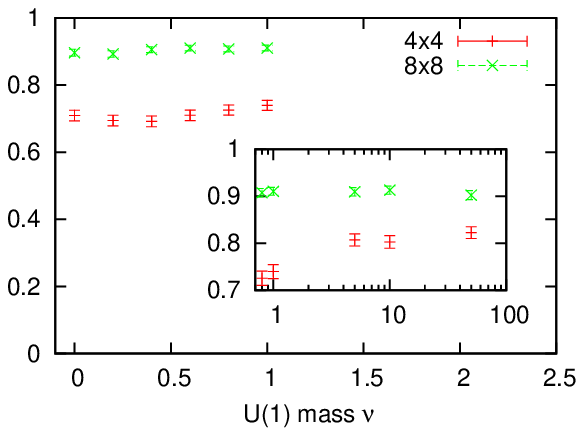}

 \caption{[CKKU] The expectation value of real part of the determinant
phase factor vs. the $U(1)$ mass $\nu$, in the $U(2)$ CKKU model.
 The $U(N)$ mass is fixed to $\mu=1.0$ and the physical volume is fixed to
 $1.0\times 1.0$. A factor $(-1)^{N-1}$ is multiplied.
}
 \label{fig:cos_u1massdep}
\end{figure}

In order to justify our treatment of the $U(1)$ mass, we have checked
the $U(1)$ mass dependence of the Wilson loop
(Fig.~\ref{fig:wilson_u1massdep}), the norm of the $SU(N)$ part of scalar
(Fig.~\ref{fig:su_scalar_u1massdep}), the distribution of the argument of
the determinant (Fig.~\ref{fig:un_det_u1massdep}) 
and its cosine (Fig.~\ref{fig:cos_u1massdep}).
One can see that there is almost no $U(1)$ mass dependence.

\begin{figure}
 \hfil
 \includegraphics[width=0.49\linewidth]{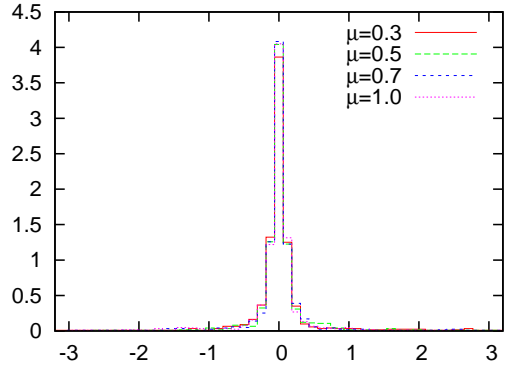}
  \caption{[CKKU] The scalar mass dependence of the argument of
 the determinant in the $U(2)$ theory. 
 The lattice size is $6\times 6$, the physical volume is fixed 
 to $0.5\times 0.5$.
 The $U(N)$ mass is varied while the $U(1)$ mass is fixed to be $\nu=5.0$.
}
 \label{fig:u2_det_L0.5_mdep}
\end{figure}

In Fig.~\ref{fig:u2_det_L0.5_mdep} we have plotted the phase distribution 
at various values of the $U(N)$ mass $\mu$, while other parameters are fixed. 
It turns out that the $\mu$-dependence is small. 

\begin{figure}
 \hfil
 \includegraphics[width=0.49\linewidth]{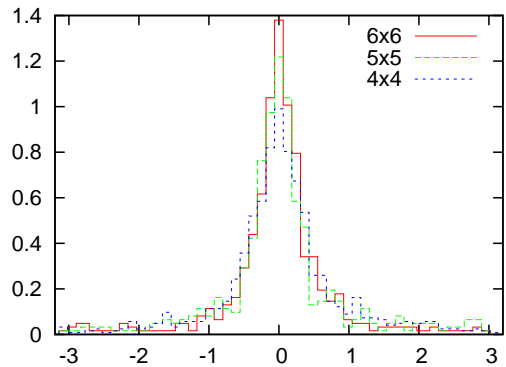}
  \caption{[CKKU] The argument of the determinant in the U(3) theory.
 The physical volume is the $L=1.00$, the $U(N)$ mass is
 $\mu=1.0$ and the $U(1)$ mass is $\nu=5.0$. }
 \label{fig:u3_det_L1.00_m1.0}
\end{figure}

In Fig.~\ref{fig:u3_det_L1.00_m1.0} we plot the phase distribution 
in the $U(3)$ theory. The distribution is broader compared to the $U(2)$ case, 
but peaks around $0$ in the continuum limit. 

\begin{figure}
 \hfil
 \includegraphics[width=0.49\linewidth]{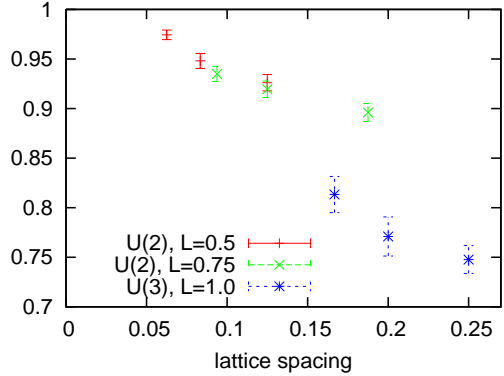}
  \caption{[CKKU] The expectation value of the real part of the determinant
 phase factor vs.\ the lattice spacing. 
  A factor $(-)^{N-1}$ is multiplied. $\mu=1.0$, $\nu=5.0$.}
 \label{fig:cos_m1.0}
\end{figure}

\begin{figure}
 \hfil
 \includegraphics[width=0.49\linewidth]{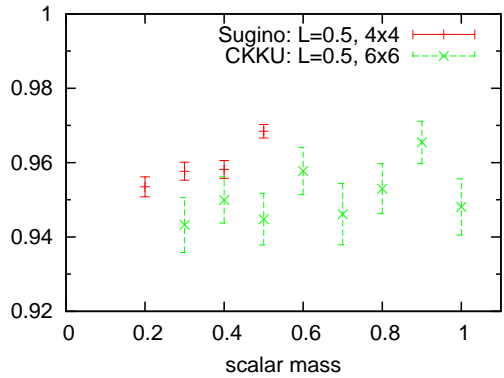}
 \caption{[CKKU and Sugino] The scalar mass $\mu$ 
dependences of the real part of the phase 
 factor in the $SU(2)$ theory (Sugino model, $4\times 4$ lattice) and 
the $U(2)$ theory (CKKU model, $6\times 6$ lattice and $\nu=5.0$).
  The physical volume is $0.5\times 0.5$ for both cases.}
 \label{fig:N2_cos_massdep}
\end{figure}

In Fig.~\ref{fig:cos_m1.0} we plot the real part of the phase.
It is clearly seen that it approaches to $1$ as lattice spacing becomes
small for each physical volume. 
The scalar mass dependence of the real part is plotted 
in Fig.~\ref{fig:N2_cos_massdep}, which shows almost no dependence.
The detailed values are listed in Tables~\ref{tab:ckku_cos1},
\ref{tab:ckku_cos2} and \ref{tab:ckku_cos3}.

\begin{table}
 \hfil
 \begin{tabular}{c|c|c|c}
  $N $ & $L$ & $ N_x=N_y $ &  Real part of the phase factor\\
 \hline
   2 & 0.50 & 4 &  0.926(8) \\
     & 0.50 & 6 &  0.948(8) \\    
     & 0.50 & 8 &  0.974(5) \\
     & 0.75 & 4 &  0.896(9) \\
     & 0.75 & 6 &  0.919(8) \\
     & 0.75 & 8 &  0.935(8) \\
     & 1.00 & 4 &  0.807(13) \\
     & 1.00 & 6 &  0.908(8) \\
     & 1.00 & 8 &  0.907(9) \\
\hline
   3 & 1.00 & 4 &  0.748(14) \\
     & 1.00 & 5 &  0.771(20) \\
     & 1.00 & 6 &  0.813(18)
 \end{tabular}
\caption{[CKKU] The real part of the phase factor in the $U(N)$ theory. 
The $U(N)$ mass and the $U(1)$ mass are fixed to $\mu=1.0$ and $\nu=0.5$,
respectively.
}
\label{tab:ckku_cos1}
\end{table}

\begin{table}
 \hfil
 \begin{tabular}{c|c|c}
  $N $ & $\mu$ &  Real part of the phase factor\\
 \hline
   2 &  0.3 &  0.943(7) \\
     &  0.4 &  0.950(6) \\    
     &  0.5 &  0.945(7) \\
     &  0.6 &  0.958(6) \\
     &  0.7 &  0.946(8) \\
     &  0.8 &  0.953(7) \\
     &  0.9 &  0.965(6) \\
     &  1.0 &  0.948(8)
 \end{tabular}
\caption{[CKKU] The real part of the phase factor in the $U(2)$ theory. 
 The physical volume is fixed to be $0.5\times 0.5$ and the lattice size 
 is $6\times 6$.  The $U(1)$ mass is fixed to $\nu=5.0$.}
\label{tab:ckku_cos2}
\end{table}

\begin{table}
 \hfil
 \begin{tabular}{c|c|c|c}
  $N $ & $\nu$ &  \multicolumn{2}{c}{Real part of the phase factor}\\
       &       &   $N_x=N_y=4$ &  $N_x=N_y=8$ \\
 \hline
   2  &   0.0  & 0.709(16)  &  0.897(10) \\
      &   0.2  & 0.694(16)  &  0.893(10) \\
      &   0.4  & 0.692(16)  &  0.906(09) \\
      &   0.6  & 0.710(16)  &  0.911(08) \\
      &   0.8  & 0.726(15)  &  0.908(09) \\
      &   1.0  & 0.740(15)  &  0.911(08) \\
      &   5.0  & 0.807(13)  &  0.910(09) \\
      &  10.0  & 0.803(13)  &  0.913(09) \\
      &  50.0  & 0.823(13)  &  0.902(10) \\
\end{tabular}
\caption{[CKKU] The real part of the phase factor in the $U(2)$ theory, 
with various values of $U(1)$ mass $\nu$.
 The physical volume is fixed to be $0.5\times 0.5$ and the lattice size 
 is $4\times 4$ and $8\times 8$.  The $U(N)$ mass is fixed to $\mu=1.0$.}
\label{tab:ckku_cos3}
\end{table}

\section{Absence of the sign problem in Sugino model}\label{sec:Sugino model}

In this section, we study the Sugino model\footnote{
See Appendix \ref{appendix:sugino_model} for the details of the model.
}
and observe the argument of the Pfaffian 
of the Dirac operator.  As before, we use the scalar mass term 
to regularize the flat direction of the potential. As shown in \cite{HK09}, 
we have checked that the scalar eigenvalues remain close enough to the origin. 
Therefore we are observing the non-abelian phase, and at the same time, 
can avoid an unphysical vacuum with large scalar eigenvalues of 
the cutoff scale. The configurations for $N=2$ is taken in this work, 
while those for 
$N=3,4,5$ are taken from the previous work \cite{HK09}.

\begin{figure}
 \hfil
 \includegraphics[width=0.49\linewidth]{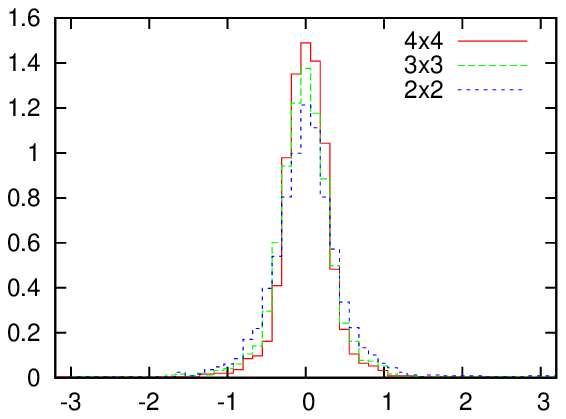}
 \hfil
 \includegraphics[width=0.49\linewidth]{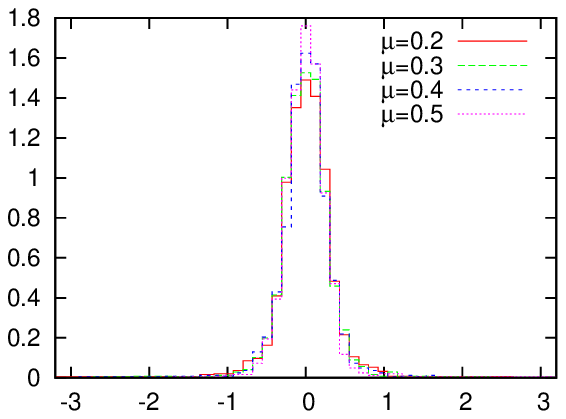}

 \caption{[Sugino] The argument of the Pfaffian for the $SU(2)$ case. 
The physical volume  is fixed to $0.5 \times 0.5$.
The left panel is with various lattice spacings and the right is with
 various scalar masses $\mu$.
}
 \label{fig:su2_pf}
\end{figure}

\begin{figure}
 \hfil
 \includegraphics[width=0.49\linewidth]{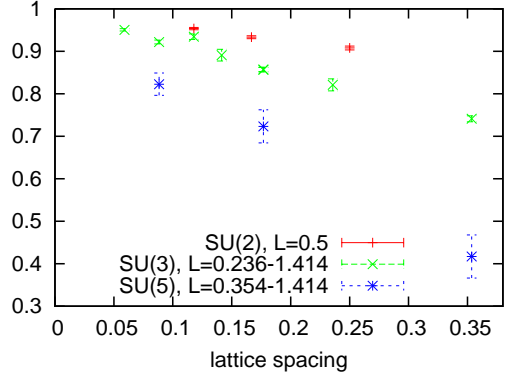}
 \caption{[Sugino] The real part of the phase factor 
 in the $SU(2)$, $SU(3)$ and $SU(5)$ theories, 
 with the physical volume $0.236\times 0.236$--$1.414\times 1.414$.
}
 \label{fig:cos}
\end{figure}

Let us start with $SU(2)$. 
In Fig.~\ref{fig:su2_pf}, we plot the distribution of 
the argument of the Pfaffian for fixed 
physical volume on the left panel.  
The peak around 0 becomes sharper as we go close to the continuum.  
On the right panel, the scalar mass dependence is plotted.
Heavier mass gives slightly shaper peak around 0, but the
mass-dependence is small. 
The average of the real part of the Pfaffian phase factor is plotted in 
Fig.~\ref{fig:cos}, which shows clear convergence to $1$ 
as the lattice spacing becomes small. (Note that $SU(2)$ case in
the plot corresponds to the continuum limit, which uses the fixed
volume.)
See also Fig.~\ref{fig:N2_cos_massdep} for the scalar mass dependence.

\begin{figure}
 \hfil
 \includegraphics[width=0.49\linewidth]{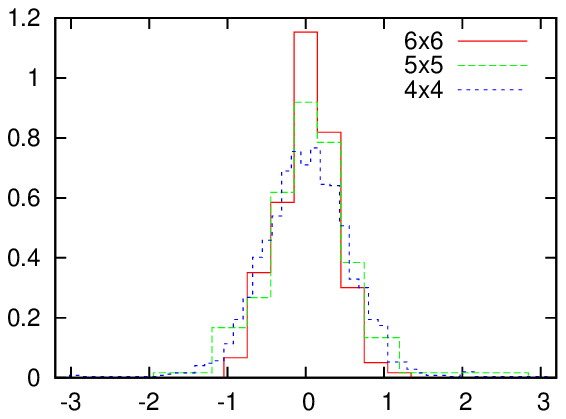}
 \hfil
 \includegraphics[width=0.49\linewidth]{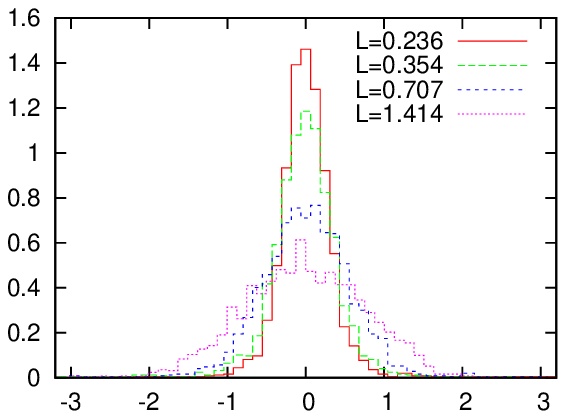}

  \caption{[Sugino] The argument of the Pfaffian in the $SU(3)$ theory. 
The scalar mass is $\mu=0.20$.
 The left panel is for a fixed volume $0.707\times 0.707$
and thus different lattice spacings.
 The right panel is for a fixed $4\times 4$ lattice with various
 physical volumes (thus various lattice spacings).
}
 \label{fig:su3_pf}
\end{figure}

In Fig.~\ref{fig:su3_pf}, we plot the lattice spacing dependence 
for the $SU(3)$ theory on the left panel. 
The peak becomes sharper as we go closer to the continuum. 
On the right panel, we show that the phase distribution 
with fixed lattice size $4\times 4$.  
The peak becomes sharper at smaller volume (or 
equivalently at smaller lattice spacing).

\begin{figure}
 \hfil
 \includegraphics[width=0.49\linewidth]{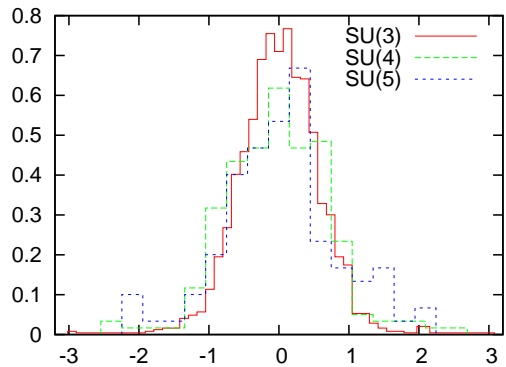}
  \caption{[Sugino] The argument of the Pfaffian 
  in the $SU(3),\ SU(4)$ and $SU(5)$ theories.  
 The lattice size is $4\times 4$ 
 and the physical volume is $0.707\times 0.707$.}
 \label{fig:pf_Ndep}
\end{figure}

The dependence on $N$ is plotted in Fig.~\ref{fig:pf_Ndep}.
As $N$ becomes large, the distribution spreads.
This can be also seen in Fig~\ref{fig:cos}, where the lattice spacing
dependence of the real part of the phase factor is plotted for $N=2$,
$3$ and $5$.  The smaller the lattice spacing is, the closer the real
part to $1$.  And larger $N$ shows slower approach to $1$.

The results are listed in terms of the real part of the phase factor
in Tables~\ref{tab:sugino_cos1} and \ref{tab:sugino_cos2}.

\begin{table}
 \hfil
 \begin{tabular}{c|c|c|c}
  $N $ & $ N_x=N_y $ & scalar mass $\mu$ & Real part of the phase factor\\
 \hline
   2 & 2 & 0.2 &  0.907(4) \\
     & 3 & 0.2 &  0.933(3) \\    
     & 4 & 0.2 &  0.953(2) \\
     & 4 & 0.3 &  0.958(2) \\
     & 4 & 0.4 &  0.958(2) \\
     & 4 & 0.5 &  0.968(2)
 \end{tabular}
\caption{[Sugino] The real part of the phase factor in the $SU(2)$ theory. 
 The physical volume is fixed to be $0.5\times 0.5$.}
 \label{tab:sugino_cos1}
\end{table}

\begin{table}
 \hfil
 \begin{tabular}{c|c|c|c}
  $N $ & $L$ & $N_x=N_y$ & Real part of the phase factor\\
 \hline
  3 & 0.236 & 4 & 0.950(3) \\
    & 0.354 & 4 & 0.922(4) \\
    & 0.707 & 4 & 0.857(5) \\
    & 0.707 & 5 & 0.891(14) \\
    & 0.707 & 6 & 0.935(7) \\
    & 1.414 & 4 & 0.742(7) \\
    & 1.414 & 6 & 0.822(14) \\
  \hline
  4 & 0.707 & 4 & 0.788(21) \\
  \hline
  5 & 0.354 & 4 & 0.82(3) \\
    & 0.707 & 4 & 0.72(4) \\
    & 1.414 & 4 & 0.42(5) \\
\end{tabular}
\caption{[Sugino] The real part of the phase factor 
in the $SU(3),\ SU(4)$  and $SU(5)$ theories. 
 The scalar mass is $\mu=0.2$.}
 \label{tab:sugino_cos2}
\end{table}

\section{Comparison of the CKKU model and the Sugino model }
\label{sec:CKKU_SUGINO_COMPARISON}
In order to confirm that our simulation captures the continuum physics, 
we compare the CKKU model and the Sugino model. In the latter, 
detailed studies have been performed;  
it correctly reproduces analytic results in continuum \cite{HK09} and also 
restoration of the full supersymmetry has been confirmed \cite{KS08sono2}. 
Here we compare the simulation result of the CKKU model with the one of
the Sugino model with periodic boundary condition for fermions.

\begin{figure}
 \hfil
 \includegraphics{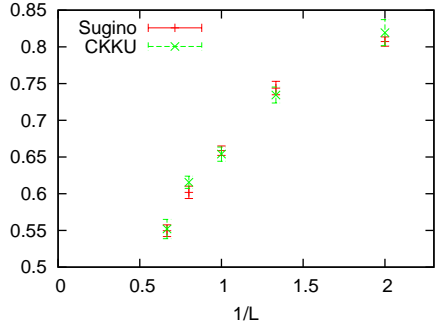}
  \caption{[CKKU and Sugino] The expectation value of the Wilson loop
 $\langle|W|\rangle$ at $\mu=1.0$ and $\nu=5.0$. 
 The extrapolation to the continuum limit has been performed.
 The gauge grope is $U(2)$ and $SU(2)$, respectively.
 }
 \label{fig:Wilson_comparison}
\end{figure}

\begin{figure}
 \hfil
 \includegraphics{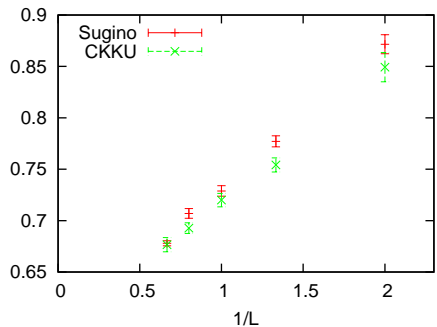}
 \caption{[CKKU and Sugino] The extent of the scalar $\sqrt{Tr(X_i(x))^2/N}$ 
(contraction w.r.t $i$ is not taken).  
The masses are $\mu=1.0$ (Sugino and CKKU) and $\nu=5.0$ (CKKU).  
 The extrapolation to the continuum limit has been performed.
 For CKKU, only the $SU(N)$ part is plotted.
 The gauge grope is $U(2)$ and $SU(2)$, respectively.
}
 \label{scalar_comparison.eps}
\end{figure}

Because the simulation of Sugino model was performed with $SU(N)$ gauge
group while for the CKKU model gauge group was chosen to be $U(N)$, we
compare the absolute value of the Wilson loop 
$W=\frac{1}{N}Tr\, e^{i\oint dx A_x}$, from which the $U(1)$ part decouples. 
As can be seen from Fig.~\ref{fig:Wilson_comparison}, 
two models give the same result in the continuum limit. 
We also compare the size of the $SU(N)$ part of the scalar fields 
(Fig.~\ref{scalar_comparison.eps}). 
Two models agree reasonably well with each other.  
Therefore we conclude 
that the both models converge to the same continuum limit as
expected.

Before concluding this section, let us comment on the flat direction. 
In~\cite{HK09} it has been shown that the flat direction is 
lifted dynamically at large-$N$, 
both in the continuum theory and in the Sugino model, and the
Sugino model converges to the correct supersymmetric continuum limit. 
In this section
we have seen the CKKU model converges to the same limit as well, and
hence we can expect the same uplift of the flat direction 
in the CKKU model.

\section{Why was a ``sign problem'' observed in previous works?}
\label{subsec:Giedt}

In \cite{Giedt03} it has been pointed out that 
the sign of the fermion determinant of $\mathcal{N}=(2,2)$ CKKU model
fluctuates violently if one chooses 
randomly generated lattice field configurations. 
However, as correctly argued in \cite{Giedt03},  
it does not mean a problem in the continuum limit -- 
randomly generated configurations are usually measure zero in 
the path integral, and hence   
it is necessary to study the distribution of the phase 
in the phase quenched simulation. If the distribution 
peaks around one (i.e. determinant is real positive) in the continuum limit, 
the sign problem does not exist. 
More crucial thing 
is the treatment of the pseudo zero-mode.
As already mentioned in \cite{Giedt03}, removing only the exact
zero-mode leads to the fluctuation of the phase factor.
The pseudo zero-mode, which should give zero eigenvalue and should 
be decoupled in the continuum limit,
gives non-zero eigenvalue due to the lattice artifact.\footnote{
As discussed in \cite{Giedt03}, one can regard this artifact
is caused by the orbifold projection, which does not commute with
manipulation needed to prove the positivity of the determinant.
}
Since what we want to extract from the lattice simulation 
is the continuum limit, we must remove a contribution from the 
pseudo zero-mode when we calculate the determinant.  
Then we obtain the correct positive determinant.

Next let us consider a result from an importance sampling 
in $\mathcal{N}=(2, 2)$ theory \cite{Catterall08, Catterall:2009it}. 
According to the plot in the paper, the distribution of the scalar eigenvalues 
is large; 
the tail of the distribution reaches to $({\rm lattice\ spacing})^{-1}$. 
However, in order for the lattice considered there 
to converge to the continuum limit \emph{at tree level}, 
the scalar eigenvalues must be of order $({\rm lattice\ spacing})^{0}$.  
Therefore it is plausible that 
the simulation does not capture the continuum physics
\footnote{ 
It has been remarked that  
large phase fluctuation arise when scalars take large expectation values
\cite{lat2010private}, 
and that this ``fluctuation of sign'' suggests the SUSY breaking because 
it can make the Witten index vanish. 
However, if such configuration corresponded to continuum theory, 
it must be an abelian phase, which does not have any dynamics, 
and hence the supersymmetry cannot be broken.  
}. 
In a simulation of the maximally supersymmetric theory 
reported in \cite{Catterall08} 
the distribution of the scalar eigenvalues is narrower compared to 
${\cal N}=(2,2)$ theory, but it is still wide (larger than the size of 
fluctuation in $4\times 4$ lattice in Fig.~\ref{fig:U2_Vol=100_M100U000scalar}). 
Therefore it is difficult to obtain robust statements for the continuum limit  
unless studying smaller lattice spacing using bigger lattice. 
Another subtlety is that in \cite{Catterall08} the system is projected to
$SU(N)$ from $U(N)$.  Although it is reported that there is
no effect to the supersymmetry in this projection, it might have
affected to the sign of determinant.

One interesting observation in \cite{Catterall08} for the maximally
supersymmetric  
theory is that the phase is close to one when the gauge group is $SU(2)$. 
The same behavior is observed also in one-dimensional theory \cite{HNSY10}. 
In this case there is no apparent kinematic reason like the pair structure 
of eigenvalues, and it is not clear if the absence of sign persists 
in the large volume and/or in the continuum limits. However, if it survives 
to some extent, it may allow detailed study of the $SU(2)$ theory 
by phase-quench or reweighting.

In ${\cal N}=(4,4)$ theory, 
it has been reported that a lattice model a la CKKU suffers
from the sign problem \cite{Giedt04}.  This 8 supercharge system is
known to have the sign problem in general so the result itself is reasonable.  
It is interesting to see whether the result changes 
when pseudo zero-mode is removed, although we do not expect the pair structure 
of eigenvalues in this theory. 
In addition, it is not clear whether the flat direction was under control in
\cite{Giedt04}, which is an important point to be studied.

\section{Conclusion}

In this paper we established the absence of the sign problem 
in 2d ${\cal N}=(2,2)$ super Yang-Mills theory on lattice. 
We studied two different lattice models: The Sugino model and
the Cohen-Kaplan-Katz-Unsal (CKKU) model. 
We have clarified the structure of the light modes 
in the CKKU model and pointed out the importance of the removal of the 
pseudo zero mode. 
We also confirmed that the both lattice models provide 
the same continuum physics as expected.  

As we pointed out in \S~\ref{subsec:Giedt}, and also 
discussed in \cite{HK09}, in order to obtain 
correct continuum limit it is crucial to control the scalar flat direction. 
From this point of view, it is possible that 
some of the past simulations for two-dimensional 
super Yang-Mills theories failed to capture the continuum physics. 
It is urgent to check whether the scalar flat direction 
was under control.

\section*{Acknowledgement}
We would like to thank O.~Aharony, L.~Mannelli and M.~\"{U}nsal 
for stimulating discussions and comments. 
A part of simulation codes was developed from the ones used in collaborations 
of I~.K. with H.~Suzuki. 
The computations were carried out partly on PC clusters at Yukawa Institute 
and RIKEN RSCC. 
The work of I.~K. was in part supported by the Nishina Memorial Foundation. 
The work of M.~H. was in part supported by JSPS Postdoctoral Fellowship
for Research Abroad. 
M.~H. would like to appreciate wonderful hospitality of Israeli people 
during three years of happy time at Weizmann Institute.  

\appendix

\section{An overall sign of the determinant of Dirac operator in
 the CKKU model}\label{appendix:eigenmodes}
Let us introduce the following notation, 
\begin{align}
 V_{1}^{ab}(\vec{n})&= \sqrt{2}\,a \tr(T_a x_{\vec{n}} T_b), \\
 \bar{V}_{1}^{ab}(\vec{n})&= \sqrt{2}\,a \tr(T_a \bar{x}_{\vec{n}} T_b), \\
 V_{2}^{ab}(\vec{n})&= \sqrt{2}\,a \tr(T_a y_{\vec{n}} T_b), \\
 \bar{V}_{2}^{ab}(\vec{n})&= \sqrt{2}\,a \tr(T_a \bar{y}_{\vec{n}} T_b),
\end{align}
where $T_a$ is a Hermitian gauge generator normalized 
as $\tr(T_a T_b)=\delta_{ab}$.
$V_\mu$ and $\bar{V}_\mu$ are related by 
\begin{equation}
 \bar{V}_\mu = (V_\mu^*)^T = V^\dagger_\mu. 
\end{equation}
The fermionic part of the action is expressed as\footnote{
A factor $N/a$ in front of $D$, which is irrelevant to the argument here,
is omitted.
} 
\begin{align}
 S_{\rm F}
  &= i(\alpha_{\vec{n}}^a, \beta_{\vec{n}}^a)
 \begin{pmatrix}
  D^{ab}_{\vec{n}\vec{m}}|_{\alpha\lambda} &   D^{ab}_{\vec{n}\vec{m}}|_{\alpha\xi} \\
  D^{ab}_{\vec{n}\vec{m}}|_{\beta\lambda} &   D^{ab}_{\vec{n}\vec{m}}|_{\beta\xi}
 \end{pmatrix}
 \begin{pmatrix}
   \lambda_{\vec{m}}^b \\ \xi_{\vec{m}}^b
 \end{pmatrix}
 = i(\alpha_{\vec{n}}^a, \beta_{\vec{n}}^a)  D^{ab}_{\vec{n}\vec{m}}
 \begin{pmatrix}
   \lambda_{\vec{m}}^b \\ \xi_{\vec{m}}^b
 \end{pmatrix}
\end{align}
with 
\begin{align}
 D^{ab}_{\vec{n}\vec{m}}|_{\alpha,\lambda}
  &=  -i\left(\delta_{\vec{n},\vec{m}} \bar{V}_{1}(\vec{n}) -\delta_{\vec{m},\vec{n}+\hat{1}}V_1^*(\vec{n})\right)^{ab},\\
 D^{ab}_{\vec{n}\vec{m}}|_{\beta,\lambda}
  &=  -i\left(\delta_{\vec{n},\vec{m}} \bar{V}_2(\vec{n}) -\delta_{\vec{m},\vec{n}+\hat{2}}V_2^*(\vec{n})\right)^{ab},\\
 D^{ab}_{\vec{n}\vec{m}}|_{\alpha,\xi}
 &= -i\left( -\delta_{\vec{n},\vec{m}} V_2(\vec{n}+\hat{1}) +\delta_{\vec{m},\vec{n}-\hat{2}}\bar{V}_2^*(\vec{n}-\hat{2})\right)^{ab},\\
 D^{ab}_{\vec{n}\vec{m}}|_{\beta,\xi}
 &= -i\left( \delta_{\vec{n},\vec{m}} V_1(\vec{n}+\hat{2}) -\delta_{\vec{m},\vec{n}-\hat{1}}\bar{V}_1^*(\vec{n}-\hat{1})\right)^{ab},
\end{align}
where each of the fermion components is defined as
\begin{equation}
 \alpha_{\vec{n}} = T_a \alpha_{\vec{n}}^a, \quad \text{etc.}
\end{equation}

Assuming the bosonic fields are constants
we obtain the momentum representation as
\begin{equation}
 D(p)=
 -i
 \begin{pmatrix}
  \bar{V}_1 -e^{iap_1} V_1^*  & -V_2 + e^{-iap_2} \bar{V}_2^* \\
  \bar{V}_2 -e^{iap_2} V_2^*  & V_1 - e^{-iap_1}\bar{V}_1^*
 \end{pmatrix}.
\end{equation}
We further decompose it into
\begin{equation}
 D(p)
  = \begin{pmatrix}
     e^{-\frac{iap_1}{2}} & 0 \\ 0 & e^{-\frac{iap_2}{2}}
    \end{pmatrix}
     D'(p)
     \begin{pmatrix}
      1 & 0 \\ 0 & e^{\frac{iap_1}{2}+\frac{iap_2}{2}}
     \end{pmatrix},
\end{equation}
where $D'$ has the same determinant as $D$. 
The explicit form of $D'$ is
\begin{align}
 D'(p)&=
 -i
 \begin{pmatrix}
  V_1\, e^{-\frac{iap_1}{2}} - V_1^*e^{\frac{iap_1}{2}} 
  & -V_2\, e^\frac{iap_2}{2} + V_2^*e^{-\frac{iap_2}{2}} \\
 V_2\, e^{-\frac{iap2_2}{2}} - V_2^*e^{\frac{iap_2}{2}}
  & V_1\, e^\frac{iap_1}{2} - V_1^* e^{-\frac{iap_1}{2}}
 \end{pmatrix}.
\end{align}
Note that because of a factor $a/2$ in the exponents, 
$D'(p)$ is not periodic w.r.t. $2\pi/a$ so that we have to
be careful about the treatment of the the boundary of the Brillouin zone.  
We use a region $-\pi/a < p_i \leq \pi/a$,  where $-\pi/a$ is not included.
It is easy to see that it satisfies
\begin{equation}
 \sigma_2 D'(p) \sigma_2 = D'^*(-p).
\end{equation}
Therefore, for $p$ which has $-p$ in our Brillouin zone, 
we have the following quartet of the eigenvalues:\footnote{
Do not confuse eigenvalue $\lambda$ with one of the fermion
or 't Hooft coupling.
}
\begin{equation}
 \lambda(p), \lambda(-p), \lambda^*(p), \lambda^*(-p)
\end{equation}
and the contribution to the determinant is always positive.
For $p=0$, we have a pair $\lambda(0), \lambda^*(0)$ which
has a positive contribution as well.
Note that for $p=0$, $D(p)=D'(p)$ thus the eigenvalues of $D(p=0)$
make a pair $(\lambda, \lambda^*)$ as well.

If the lattice is $\text{(odd)}\times\text{(odd)}$, we do not have
$p_i=\pi/a$ modes so the determinant of $D'$ and thus that of $D$
is positive.

If the lattice is $\text{(even)}\times\text{(even)}$, since
we have $p_i=\pi/a$ modes and  $p_i=-\pi/a$ is not in the Brillouin zone,
the sign becomes non-trivial.
For $p_1=\pi/a$, we have
\begin{equation}
 \sigma_2 D'(\pi/a,p_2) \sigma_2 = D'(\pi/a,-p_2)^*
\end{equation}
for $p_2\neq \pi/2$.
Therefore we have a quartet
$(\lambda(p_2), \lambda(-p_2), \lambda(p_2)^*, \lambda^*(-p_2) )$
for $p_2 \neq 0$ and a doublet $(\lambda, \lambda^*)$ for $p_2=0$,
both of them give positive contributions to the determinant.
The situation is the same when $p_2=\pi/a$.
The remaining combination is $p=(\pi/a, \pi/a)$.
This time we have
\begin{equation}
 D'(p_1=\pi/a, p_2=\pi/a)
 = -\begin{pmatrix}
     \bar{V}_1 +V_1^* & V_2 + \bar{V}_2^* \\
     \bar{V}_2 +V_2^* & -V_1-\bar{V}_1^*
    \end{pmatrix}
\end{equation}
which satisfies $\sigma_2 D' \sigma_2 = -D'^*$.
The eigenvalues make a pair $(\lambda, -\lambda^*)$.
The sign contribution to the determinant from this sector
is $(-1)^{N^2}=(-1)^N$. 

For $iD$, an extra factor 
$i^{2N^2 \times \text{(num. of lattice sites)}}$ appears, 
and hence the phase is $(-1)^N$ for both ${\rm odd}\times {\rm odd}$ 
and ${\rm even}\times {\rm even}$ lattices. 
Since we remove the two lightest modes 
--- one exact zero mode and one pseudo zero mode --- the determinant of 
$iD$ picks up an extra factor $i^{-2}$ and thus the determinant of $iD$
has a sign $(-1)^{N-1}$.

\section{The action of the Sugino model}
\label{appendix:sugino_model}
Sugino's lattice action \cite{Sugino04} is given by 
\footnote{Here we follow the notation in \cite{Suzuki07}
with a slightly different normalization.  
Although group theoretical normalizations are different
in \cite{Suzuki07} and \cite{CKKU03}, one can absorb them
by rescaling the 't Hooft couplings as
$2\lambda_{\rm CKKU}=\lambda_{\rm Sugino}$.  We set
$\lambda=\lambda_{\rm CKKU}=1$.
} 
\begin{eqnarray}
S_{lattice}
=
a_x a_y\sum_{\vec{x}}
\left\{
\sum_{i=1}^3{\cal L}_{Bi}(\vec{x})
+
\sum_{i=1}^6{\cal L}_{Fi}(\vec{x})
\right\}
+{\rm (auxiliary\ field)}, 
\end{eqnarray}
where 
\begin{eqnarray}
{\cal L}_{B1}(\vec{x})
&=&
\frac{N}{8a_x^2 a_y^2}
Tr
[\phi(\vec{x}),\bar{\phi}(\vec{x})]^2, 
\\
{\cal L}_{B2}(\vec{x})
&=&
\frac{N}{8a_x^2 a_y^2}
Tr
\hat{\Phi}_{TL}(\vec{x})^2, 
\\
{\cal L}_{B3}(\vec{x})
&=&
\frac{N}{2a_x^3 a_y}
Tr\bigl\{
\left(
\phi(\vec{x})-U_x(\vec{x})\phi(\vec{x}+a_x\hat{x})U_x(\vec{x})^{-1}
\right)
\nonumber\\
& &
\hspace{3cm}\times
\left(
\bar{\phi}(\vec{x})-U_x(\vec{x})\bar{\phi}(\vec{x}+a_x\hat{x})U_x(\vec{x})^{-1}
\right)
\bigl\}
\nonumber\\
& &
+
\frac{N}{2a_x a_y^3}
Tr\bigl\{
\left(
\phi(\vec{x})-U_y(\vec{x})\phi(\vec{x}+a_y\hat{y})U_y(\vec{x})^{-1}
\right)
\nonumber\\
& &
\hspace{3cm}\times
\left(
\bar{\phi}(\vec{x})-U_y(\vec{x})\bar{\phi}(\vec{x}+a_y\hat{y})U_y(\vec{x})^{-1}
\right)
\bigl\}
\end{eqnarray}
and
\begin{eqnarray}
{\cal L}_{F1}(\vec{x})
&=&
-\frac{N}{8a_x^2 a_y^2}
Tr\left(
\eta(\vec{x})[\phi(\vec{x}),\eta(\vec{x})]
\right), 
\\
{\cal L}_{F2}(\vec{x})
&=&
-\frac{N}{2a_x^2 a_y^2}
Tr\left(
\chi(\vec{x})[\phi(\vec{x}),\chi(\vec{x})]
\right), 
\\
{\cal L}_{F3}(\vec{x})
&=&
-\frac{N}{2a_x^3 a_y}
Tr\left\{
\psi_0(\vec{x})\psi_0(\vec{x})
\left(
\bar{\phi}(\vec{x})+U_x(\vec{x})\bar{\phi}(\vec{x}+a_x\hat{x})U_x(\vec{x})^{-1}
\right)
\right\}, 
\\
{\cal L}_{F4}(\vec{x})
&=&
-\frac{N}{2a_x a_y^3}
Tr\left\{
\psi_1(\vec{x})\psi_1(\vec{x})
\left(
\bar{\phi}(\vec{x})+U_y(\vec{x})\bar{\phi}(\vec{x}+a_y\hat{y})U_y(\vec{x})^{-1}
\right)
\right\}, 
\\
{\cal L}_{F5}(\vec{x})
&=&
i\frac{N}{2a_x^2 a_y^2}
Tr\left(
\chi(\vec{x})\cdot Q\hat{\Phi}(\vec{x})
\right), 
\\
{\cal L}_{F6}(\vec{x})
&=&
-i\frac{N}{2a_x^3 a_y}
Tr\bigl\{
\psi_0
\left(
\eta(\vec{x})-U_x(\vec{x})\eta(\vec{x}+a_x\hat{x})U_x(\vec{x})^{-1}
\right)
\bigl\}
\nonumber\\
& &
-i\frac{N}{a_x a_y^3}
Tr\bigl\{
\psi_1
\left(
\eta(\vec{x})-U_y(\vec{x})\eta(\vec{x}+a_y\hat{y})U_y(\vec{x})^{-1}
\right)
\bigl\}, 
\end{eqnarray}
where $U(\vec{x},\mu)$ are gauge link variables, $\phi(\vec{x})$ is a complex scalar, 
$\eta(\vec{x})$, $\chi(\vec{x})$ and $\psi_\mu(\vec{x})$ are fermion field, 
$a_x$ and $a_y$ are lattice spacings \footnote{In the actual simulation 
we have used the isotropic lattice, $a_x=a_y$.}, 
$\epsilon$ is a real parameter which must be chosen appropriately 
for each $N$, 
\begin{eqnarray}
\hat{\Phi}(\vec{x})
&=&
\frac{-i(P(\vec{x})-P(\vec{x})^{-1})}{1-|1-P(\vec{x})|^2/\epsilon^2}, 
\qquad
\hat{\Phi}_{TL}(\vec{x})
=
\hat{\Phi}(\vec{x})
-
\frac{1}{N}\left(Tr\hat{\Phi}(\vec{x})\right)\cdot\textbf{1}, 
\end{eqnarray}
where $P(\vec{x})=U_x(\vec{x})U_y(\vec{x}+\hat{x})U_x^\dagger(\vec{x}+\hat{y})U_y^\dagger(\vec{x})$ is 
the plaquette variable, 
and $Q$ generates one of the four super transformations,
\begin{align}
 QU_\mu(\vec{x}) &= i \psi_\mu(\vec{x})U_\mu(\vec{x}), \\
 Q\psi_\mu(\vec{x})
  &= i\psi_\mu(\vec{x})\psi_\mu(\vec{x})
     -i\bigl( \phi(\vec{x})-U_\mu(\vec{x})\phi(\vec{x}+a_\mu\hat{\mu})U_\mu(\vec{x})^{-1}\bigr), \\
 Q\phi(\vec{x}) &= 0, \\
 Q\chi(\vec{x}) &= H(\vec{x}), \\
 QH(\vec{\vec{x}}) &= [\phi(\vec{x}), \chi(\vec{x})], \\
 Q\phibar(\vec{x}) &= \eta(\vec{x}), \\
 Q\eta(\vec{x}) &= [\phi(\vec{x}), \phibar(\vec{x})].
\end{align}
Sugino's action $S_{lattice}$ is invariant under the supersymmetry 
generated by $Q$, 
because $Q$ is nilpotent up to commutators 
and $S$ can be written in a $Q$-exact form. 

In \cite{Sugino04}, using super-renormalizability and symmetry argument, 
it was shown that other three supersymmetries, 
which is broken by a lattice artifact at the discretized level, 
is restored in the continuum limit.   
Furthermore, in \cite{KS08sono2}, 
this restoration has been confirmed explicitly 
by the Monte-Carlo simulation.  
Absence of operator mixing/renormalization is has been shown
perturbatively in \cite{Kadoh:2009rw}.



\begin{thebibliography}{99}

\bibitem{Martin97}
  S.~P.~Martin,
  ``A Supersymmetry Primer,''
  arXiv:hep-ph/9709356.

\bibitem{BFSS96}
T.~Banks, W.~Fischler, S.~H.~Shenker and L.~Susskind,
{ ``M theory as a matrix model: A conjecture,''}
Phys.\ Rev.\  D {\bf 55} (1997) 5112,
[arXiv:hep-th/9610043].


\bibitem{IKKT96}
N.~Ishibashi, H.~Kawai, Y.~Kitazawa and A.~Tsuchiya,
{ ``A large-N reduced model as superstring,''}
Nucl.\ Phys.\  B {\bf 498} (1997) 467,
[arXiv:hep-th/9612115].


\bibitem{DVV97}
  L.~Motl,
   ``Proposals on nonperturbative superstring interactions,'' 
  arXiv:hep-th/9701025.

R.~Dijkgraaf, E.~P.~Verlinde and H.~L.~Verlinde,
{ ``Matrix string theory,''}
Nucl.\ Phys.\  B {\bf 500} (1997) 43,
[arXiv:hep-th/9703030].


\bibitem{Maldacena97}
J.~M.~Maldacena,
{ ``The large N limit of superconformal field theories and supergravity,''}
Adv.\ Theor.\ Math.\ Phys.\  {\bf 2} (1998) 231
[Int.\ J.\ Theor.\ Phys.\  {\bf 38} (1999) 1113]
[arXiv:hep-th/9711200].


\bibitem{IMSY98}
  N.~Itzhaki, J.~M.~Maldacena, J.~Sonnenschein and S.~Yankielowicz,
  { ``Supergravity and the large N limit of theories with sixteen
   supercharges,''} 
  Phys.\ Rev.\  D {\bf 58} (1998) 046004,
  [arXiv:hep-th/9802042].


\bibitem{CKKU03} 
 A.~G.~Cohen, D.~B.~Kaplan, E.~Katz and M.~Unsal,
  ``Supersymmetry on a Euclidean spacetime lattice. I: A target theory with
  four supercharges,''
  JHEP {\bf 0308} (2003) 024, 
  [arXiv:hep-lat/0302017].

\bibitem{Sugino04}
  F.~Sugino,
  ``Super Yang-Mills theories on the two-dimensional lattice with exact
  supersymmetry,''
  JHEP {\bf 0403} (2004) 067,
  [arXiv:hep-lat/0401017].
 

\bibitem{2d other formulations} 
  A.~G.~Cohen, D.~B.~Kaplan, E.~Katz and M.~Unsal,
  ``Supersymmetry on a Euclidean spacetime lattice. II: Target theories  with
  eight supercharges,''
  JHEP {\bf 0312} (2003) 031,
  [arXiv:hep-lat/0307012].


   S.~Catterall,
  ``A geometrical approach to N = 2 super Yang-Mills theory on the two
  dimensional lattice,''
  JHEP {\bf 0411} (2004) 006,
  [arXiv:hep-lat/0410052].
  
  H.~Suzuki and Y.~Taniguchi,
  ``Two-dimensional N = (2,2) super Yang-Mills theory on the lattice via
  dimensional reduction,''
  JHEP {\bf 0510} (2005) 082,
  [arXiv:hep-lat/0507019].

\bibitem{KU_16SUSY}
   D.~B.~Kaplan and M.~Unsal,
  ``A Euclidean lattice construction of supersymmetric Yang-Mills theories
  with sixteen supercharges,''
  JHEP {\bf 0509} (2005) 042,
  [arXiv:hep-lat/0503039]. 
 
\bibitem{D'Adda:2005zk}
  A.~D'Adda, I.~Kanamori, N.~Kawamoto and K.~Nagata,
  ``Exact extended supersymmetry on a lattice: Twisted N = 2 super  Yang-Mills
  in two dimensions,''
  Phys.\ Lett.\  B {\bf 633} (2006) 645,
  [arXiv:hep-lat/0507029].
 

\bibitem{Catterall:2009it}
  S.~Catterall, D.~B.~Kaplan and M.~Unsal,
  ``Exact lattice supersymmetry,''
  Phys.\ Rept.\  {\bf 484}, 71 (2009)
  [arXiv:0903.4881 [hep-lat]].



\bibitem{4d-n1-SYM}
  J.~Giedt, R.~Brower, S.~Catterall, G.~T.~Fleming and P.~Vranas,
  ``Lattice super-Yang-Mills using domain wall fermions in the chiral limit,''
  Phys.\ Rev.\  D {\bf 79}, 025015 (2009)
  [arXiv:0810.5746 [hep-lat]].

  M.~G.~Endres,
  ``Dynamical simulation of N=1 supersymmetric Yang-Mills theory with domain
  wall fermions,''
  Phys.\ Rev.\  D {\bf 79}, 094503 (2009)
  [arXiv:0902.4267 [hep-lat]].

  K.~Demmouche, F.~Farchioni, A.~Ferling, I.~Montvay, G.~Munster, E.~E.~Scholz and J.~Wuilloud,
  ``Simulation of 4d N=1 supersymmetric Yang-Mills theory with Symanzik
  improved gauge action and stout smearing,''
  arXiv:1003.2073 [hep-lat].

\bibitem{Maldacena:2002rb}
  J.~M.~Maldacena, M.~M.~Sheikh-Jabbari and M.~Van Raamsdonk,
  ``Transverse fivebranes in matrix theory,''
  JHEP {\bf 0301} (2003) 038, 
  [arXiv:hep-th/0211139].

\bibitem{HMS10}
  M.~Hanada, S.~Matsuura and F.~Sugino,
  ``Two-dimensional lattice for four-dimensional N=4 supersymmetric
  Yang-Mills,''
  arXiv:1004.5513 [hep-lat].
  
  M.~Hanada,
  ``A proposal of a fine tuning free formulation of 4d N=4 super Yang-Mills,''
  to appear in JHEP, arXiv:1009.0901 [hep-lat].

\bibitem{Ishii:2008ib}
  T.~Ishii, G.~Ishiki, S.~Shimasaki and A.~Tsuchiya,
  ``N=4 Super Yang-Mills from the Plane Wave Matrix Model,''
  Phys.\ Rev.\  D {\bf 78} (2008) 106001,
  [arXiv:0807.2352 [hep-th]].

  G.~Ishiki, S.~W.~Kim, J.~Nishimura and A.~Tsuchiya,
  ``Deconfinement phase transition in N=4 super Yang-Mills theory on $R\times S^3$ from
  supersymmetric matrix quantum mechanics,''
  Phys.\ Rev.\ Lett.\  {\bf 102} (2009) 111601, 
  [arXiv:0810.2884 [hep-th]].

  G.~Ishiki, S.~W.~Kim, J.~Nishimura and A.~Tsuchiya,
  ``Testing a novel large-N reduction for N=4 super Yang-Mills theory on
  $R\times S^3$,''
  JHEP {\bf 0909} (2009) 029, 
  [arXiv:0907.1488 [hep-th]].

\bibitem{Hanada:2009hd}
  M.~Hanada, L.~Mannelli and Y.~Matsuo,
  ``Large-N reduced models of supersymmetric quiver, Chern-Simons gauge
  theories and ABJM,''
  JHEP {\bf 0911} (2009) 087, 
  [arXiv:0907.4937 [hep-th]].


  G.~Ishiki, S.~Shimasaki and A.~Tsuchiya,
  ``A Novel Large-N Reduction on $S^3$: Demonstration in Chern-Simons Theory,''
  Nucl.\ Phys.\  B {\bf 834} (2010) 423,
  [arXiv:1001.4917 [hep-th]].



\bibitem{HNT07}
  M.~Hanada, J.~Nishimura and S.~Takeuchi,
  { ``Non-lattice simulation for supersymmetric gauge theories in one
   dimension,''} 
  Phys.\ Rev.\ Lett.\  {\bf 99} (2007) 161602,
  [arXiv:0706.1647 [hep-lat]].
  
  
\bibitem{AHNT07} 
  K.~N.~Anagnostopoulos, M.~Hanada, J.~Nishimura and S.~Takeuchi,
  { ``Monte Carlo studies of supersymmetric matrix quantum mechanics with sixteen
   supercharges at finite temperature,''} 
   Phys.\ Rev.\ Lett.\  {\bf 100} (2008) 021601,
  [arXiv:0707.4454 [hep-th]].


\bibitem{HMNT08}
  M.~Hanada, A.~Miwa, J.~Nishimura and S.~Takeuchi,
   ``Schwarzschild radius from Monte Carlo calculation of the Wilson loop in
   supersymmetric matrix quantum mechanics,''
    Phys.\ Rev.\ Lett.\  {\bf 102} (2009) 181602,
   [arXiv:0811.2081 [hep-th]].
 
\bibitem{HHNT08} 
  M.~Hanada, Y.~Hyakutake, J.~Nishimura and S.~Takeuchi,
  ``Higher derivative corrections to black hole thermodynamics from
  supersymmetric matrix quantum mechanics,''
  Phys.\ Rev.\ Lett.\ {\bf 102}  (2009) 191602,
  [arXiv:0811.3102 [hep-th]].

\bibitem{HNSY09}
  M.~Hanada, J.~Nishimura, Y.~Sekino and T.~Yoneya,
  ``Monte Carlo studies of Matrix theory correlation functions,''
  Phys.\ Rev.\ Lett.\  {\bf 104} (2010) 151601,
  [arXiv:0911.1623 [hep-th]].

\bibitem{HNSY10}
  M.~Hanada, J.~Nishimura, Y.~Sekino and T.~Yoneya, 
  in preparation. 


\bibitem{CW07}
  S.~Catterall and T.~Wiseman,
  {``Towards lattice simulation of the gauge theory duals to black holes and hot
   strings,''} 
  JHEP {\bf 0712} (2007) 104,
  [arXiv:0706.3518 [hep-lat]].
  
  
  S.~Catterall and T.~Wiseman,
  ``Black hole thermodynamics from simulations of lattice Yang-Mills theory,''
  Phys.\ Rev.\  D {\bf 78} (2008) 041502,
  [arXiv:0803.4273 [hep-th]].
  

  S.~Catterall and T.~Wiseman,
  ``Extracting black hole physics from the lattice,''
  JHEP {\bf 1004} (2010) 077,
  [arXiv:0909.4947 [hep-th]].

\bibitem{HK09} 
   M.~Hanada and I.~Kanamori,
  ``Lattice study of two-dimensional N=(2,2) super Yang-Mills at large-N,''
  Phys.\ Rev.\  D {\bf 80} (2009) 065014,
  [arXiv:0907.4966 [hep-lat]].


\bibitem{Catterall:2010fx}
  S.~Catterall, A.~Joseph and T.~Wiseman,
  ``Thermal phases of D1-branes on a circle from lattice super Yang-Mills,''
  JHEP {\bf 1012} (2010) 022
  [arXiv:1008.4964 [hep-th]].

\bibitem{Catterall:2010ya}
  S.~Catterall, A.~Joseph and T.~Wiseman,
  ``Gauge theory duals of black hole - black string transitions of
  gravitational theories on a circle,''
  arXiv:1009.0529 [hep-th].

  
\bibitem{KNS98}
  W.~Krauth, H.~Nicolai and M.~Staudacher,
  ``Monte Carlo approach to M-theory,''
  Phys.\ Lett.\  B {\bf 431} (1998) 31,
  [arXiv:hep-th/9803117].
    
\bibitem{Kanamori:2010gw}
  I.~Kanamori,
  ``A Method for Measuring the Witten Index Using Lattice Simulation,''
  Nucl.\ Phys.\  B {\bf 841} (2010) 426,
  [arXiv:1006.2468 [hep-lat]].

\bibitem{AABHN00}
  J.~Ambjorn, K.~N.~Anagnostopoulos, W.~Bietenholz, T.~Hotta and J.~Nishimura,
  ``Large N dynamics of dimensionally reduced 4D SU(N) super Yang-Mills
  theory,''
  JHEP {\bf 0007} (2000) 013,
  [arXiv:hep-th/0003208].  
  
\bibitem{HMM09}
  M.~Hanada, L.~Mannelli and Y.~Matsuo,
  ``Four-dimensional N=1 super Yang-Mills from matrix model,''
  Phys.\ Rev.\  D {\bf 80}  (2009) 125001,
  [arXiv:0905.2995 [hep-th]].


\bibitem{Suzuki07}  
  H.~Suzuki,
  { ``Two-dimensional $\mathcal{N}=(2,2)$ super Yang-Mills theory on computer,''} 
  JHEP {\bf 0709}  (2007) 052,
  [arXiv:0706.1392 [hep-lat]].

\bibitem{Kanamori09}
  I.~Kanamori,
  ``Vacuum energy of two-dimensional N=(2,2) super Yang-Mills theory,''
  Phys.\ Rev.\  D {\bf 79} (2009) 115015,
  [arXiv:0902.2876 [hep-lat]].


\bibitem{Azeyanagi:2009zf}
  T.~Azeyanagi, M.~Hanada, T.~Hirata and H.~Shimada,
  ``On the shape of a D-brane bound state and its topology change,''
  JHEP {\bf 0903} (2009) 121,
  [arXiv:0901.4073 [hep-th]].
 
 
 
\bibitem{KSS07}
  I.~Kanamori, H.~Suzuki and F.~Sugino,
  {``Euclidean lattice simulation for the dynamical supersymmetry breaking,''} 
 Phys.\ Rev.\  D {\bf 77} (2008) 091502,
  [arXiv:0711.2099 [hep-lat]].
   
 I.~Kanamori, F.~Sugino and H.~Suzuki,
  {``Observing dynamical supersymmetry breaking with euclidean lattice
   simulations,''} 
  Prog.\ Theor.\ Phys.\  {\bf 119} (2008) 797,
  [arXiv:0711.2132 [hep-lat]].


\bibitem{KS08}
  I.~Kanamori and H.~Suzuki,
  ``Some physics of the two-dimensional $\mathcal{N}=(2,2)$ supersymmetric
   Yang-Mills theory: Lattice Monte Carlo study,''
  Phys.\ Lett.\  B {\bf 672}  (2009) 307,
  [arXiv:0811.2851 [hep-lat]]. 


\bibitem{Giedt03}
  J.~Giedt,
  ``Non-positive fermion determinants in lattice supersymmetry,''
  Nucl.\ Phys.\  B {\bf 668} (2003) 138,
  [arXiv:hep-lat/0304006].
 

 \bibitem{Catterall08}
  S.~Catterall,
  ``First results from simulations of supersymmetric lattices,''
  JHEP {\bf 0901} (2009) 040, 
  [arXiv:0811.1203 [hep-lat]].
 
\bibitem{Unsal:2005yh}
  M.~Unsal,
  ``Compact gauge fields for supersymmetric lattices,''
  JHEP {\bf 0511} (2005) 013, 
  [arXiv:hep-lat/0504016].

 
\bibitem{Gregory:1993vy}
  R.~Gregory and R.~Laflamme,
  ``Black strings and p-branes are unstable,''
  Phys.\ Rev.\ Lett.\  {\bf 70}  (1993) 2837,
  [arXiv:hep-th/9301052].

 
 
\bibitem{Aharony:2004ig}
  O.~Aharony, J.~Marsano, S.~Minwalla and T.~Wiseman,
  ``Black hole-black string phase transitions in thermal 1+1-dimensional
  supersymmetric Yang-Mills theory on a circle,''
  Class.\ Quant.\ Grav.\  {\bf 21}  (2004) 5169,
  [arXiv:hep-th/0406210].
 
\bibitem{KS08sono2}
  I.~Kanamori and H.~Suzuki,
  ``Restoration of supersymmetry on the lattice: Two-dimensional
  $\mathcal{N}=(2,2)$ supersymmetric Yang-Mills theory,''
  Nucl.\ Phys.\  B {\bf 811} (2009) 420,
  [arXiv:0809.2856 [hep-lat]]. 

\bibitem{lat2010private}
  S.~Catterall, a private communication.

 
\bibitem{Giedt04}
  J.~Giedt,
  ``Deconstruction, 2d lattice super-Yang-Mills, and the dynamical lattice
  spacing,''
  arXiv:hep-lat/0405021. 
 
\bibitem{Kadoh:2009rw}
  D.~Kadoh and H.~Suzuki,
  ``SUSY WT identity in a lattice formulation of 2D $\mathcal{N}=(2,2)$ SYM,''
  Phys.\ Lett.\  B {\bf 682} (2010) 466,
  [arXiv:0908.2274 [hep-lat]].
 
 
 

\end{thebibliography}
\end{document}